\documentstyle[11pt,aaspp4]{article}
\slugcomment{Submitted to The Astronomical Journal}
\lefthead{Fomalont et al.}
\righthead{Pulsar Motions Using In-Beam Calibrators}

\begin{document}

\title{Sub-Milliarcsecond Precision of Pulsar Motions:\\
    Using In-Beam Calibrators with the VLBA}

\author{E. B. Fomalont}
\affil{National Radio Astronomy Observatory, Charlottesville,
    VA 22903}

\author{W. M. Goss, and A. J. Beasley}
\affil{National Radio Astronomy Observatory, Socorro, NM  87801}

\author{S. Chatterjee}
\affil{Astronomy and Space Sciences, Cornell University, Ithaca, NY 14853}
\affil{National Radio Astronomy Observatory, Socorro, NM  87801}
\begin{abstract}
We present Very Long Baseline Array phase-referenced measurements of
the parallax and proper motion of two pulsars, B0919+06 and B1857--26.
Sub-milliarcsecond positional accuracy was obtained by simultaneously
observing a weak calibrator source within the $40'$ field of view of
the VLBA at 1.5 GHz. We discuss the merits of using weak close
calibrator sources for VLBI observations at low frequencies, and
outline a method of observation and data reduction for these type of
measurements.  For the pulsar B1919+06 we measure a parallax of
$0.31\pm 0.14$ mas.  The accuracy of the proper motions is $\approx
0.5$ mas, an order of magnitude improvement over most previous
determinations.

\end{abstract}

\keywords{astrometry---pulsars: general}

\section{Introduction}

The determination of pulsar proper motions and parallaxes using
position measurements at the sub-milliarcsecond level over many years
is relevant for many astrophysical questions: (1) The proper motion
may indicate the birth area of the pulsar and its ejection velocity
(egs. \cite{kas96}) ; (2) The unambiguous distance determined from the
parallax can determine the intrinsic properties of the pulsar and
calibrate dispersion-based distance measurements (eg. \cite{tay93});
(3) Comparison of pulsar positions derived from Very Long Baseline
Interferometry (VLBI) and that determined from pulsar timing analysis
can be used to compare the quasar reference and planetary ephemerides
(\cite{fom84}).

Since the opening of the Very Long Baseline Array (VLBA) in 1994, we
have conducted an experimental program to investigate methods for
obtaining high precision pulsar motions and parallaxes.  With the use
of VLBI techniques, radio images are routinely made with a resolution
of a few milliarcseconds.  If the object being imaged is relatively
bright and small in angular extent, positional accuracies of $\sim
1$\% of the resolution are possible.  However, because pulsars are
generally much stronger at lower frequencies, most observations are
made below 3 GHz where ionospheric refraction is large and variable,
leading to position errors and image distortion.  The primary goal of
our program was to characterize these ionospheric errors and minimize
their effect with appropriate observational and reduction techniques.

Most previous VLBI astrometric observations of pulsars used the
measured group delay and rates to derive accurate positions
(\cite{gwi86}). This technique does not use the interferometer phase
information directly, but rather the rate of change of phase with
frequency (group delay) and time (delay rate).  Although this
technique does not require long term phase stability, the accuracy is
proportional to the spanned bandwidth divided by the observing
frequency, generally less than 20\% compared with using the
interferometer phase directly.

In order to obtain sub-milliarcsecond positional accuracy for pulsars,
especially the weaker pulsars, phase connection (also known as
phase-referencing) techniques must be used.  With these techniques the
position of a pulsar is measured with respect to an adjacent celestial
source (calibrator) by alternating observations between the pulsar and
calibrator (which we will call the {\it nodding} calibrator) every few
minutes. Interpolation of the phase corrections derived for the
calibrator source to the target source removes first order effects of
instrumental and electronic delays, and unknown atmospheric,
ionospheric and geometric errors (\cite{bea95}).  As long as the
calibrator source is stationary and unchanging, it provides a firm
fiducial point over time for measuring the relative position of the
pulsar position.

However, the temporal and spatial properties of this phase connection
process---source/calibrator angular separation, temporal switching
cycle, frequency coverage, multi-calibration sources---may limit the
positional sensitivity of the technique, rather than that imposed by
the noise limits of the observations.  Since most previous work has
focused on observations at frequencies higher than 3 GHz (eg
\cite{bea95}) where tropospheric effects dominate, additional tests
were needed to examine the temporal and spatial properties of
ionospheric refraction.  Below a frequency of about 2 GHz,
differential ionospheric delays between the calibrator and target
sources may be large even for very fast switching times and small
separations.

This paper discusses one particular solution to this problem -- the
calibration of a pulsar position using a faint radio source which is
within the primary beam of the VLBA antenna---with recommendations on
procedures to obtain high precision positional accuracy. The
advantages of using in-beam calibrators are twofold: (1) there is no
repointing of the antenna required (only recorrelation at the
calibrator position), and (2) the angular separation of the target and
calibrator (e.g. $<25'$) minimizes the errors due to spatial
variations in ionospheric delay.  This technique is not limited to
pulsars, of course, but is only applicable at relatively low
frequencies where the chance of finding a suitable in-beam calibrator
is reasonable.

\section{The Selection of Pulsars and the Radio Observations}

    We selected four pulsars which were previously observed with a VLA
pulsar astrometric program between 1984 and 1993 (\cite{fom92,fom96}).
These pulsars were generally strong enough to be detected with the
VLBA without pulsar gating and all had at least one nearby background
source within the 25\,m antenna primary beam region ($<25'$) from the pulsar.
Subsequent VLA observations at higher frequency identified those background
sources with a flat radio spectrum and angular sizes less than $2''$.
Information on these four pulsars, their nodding calibrator and the
possible in-beam calibrator, is given in Table 1.

     VLBA observations were made on November 9, 1994, September 23,
1995 and April 1, 1996, each day for 16 hours.  These dates were
chosen to maximize the parallax offset in right ascension for the
sources.  In order to obtain relatively long periods of phase
connection data on each pulsar, we observed each pulsar and its
nodding calibrator for one contiguous hour, alternating hourly among
the four pulsar fields.  All observation cycles were five minutes on
pulsar and two minutes on calibrator, with about 30 seconds lost in
slewing between two observations.  This cycle time was considered to
be short enough to allow interpolation of atmospheric and ionospheric
phase changes between calibrator observations most of the time.

    The eight independent frequency channels (called IF's) were tuned
to 1410, 1418, 1442, 1586, 1583, 1642, 1678, 1694 MHz, each with 8 MHz
bandwidth, in order to span a large frequency range.  It is possible
to remove ionospheric refraction effects by comparing the images
obtained at different frequencies, although this calibration was not
needed with the in-beam calibration approach we will discuss in this
paper.  The pulsar data were correlated twice, at the pulsar position,
and at the position of the in-beam calibrator.  The data were sampled
every 2 seconds and 32 frequency channels were provided for each of
the 8 observing frequencies.

    Two additional observations of B0919+06 were made on March 26 and
March 30, 1998, as part of a larger project not originally intended
for use with this in-beam project.  The B0919+06 data were
recorrelated at the in-beam calibrator
position (J0923+068).  This additional fourth and fifth epoch (only
four days apart) of this pulsar can be used to determine possible
systematic error which can not be obtained with only three epochs.
The nodding calibrator used for these observations was not 0906+015,
but J0914+0245 (\cite {bea98}); however, the nodding calibrator is
used only to determine the gross calibration of the observations and
not for phase connection.

\section{The Data Reduction}

    The first part of the calibration of these data is identical to
that used for typical nodding calibration as practiced at the VLBA.
These procedures are summarized below.  The second part, using the
weak in-beam calibration to improve further the phase calibration, is
described in more detail.  While this calibration method has been used
and discussed previously (egs \cite{mar94}, \cite{bra99}), we are
attempting to push this reduction method to faint levels and this
requires somewhat different considerations in this part of the phase
calibrations.

\subsection{Phase Connection to the Nodding Calibrator}

    The data reduction steps used for phase-connection between a
nodding calibrator and a target source have been described by
\cite{bea95}.  The separation of these calibrators from the pulsar
fields were in the range 4 to 11 degrees and are now considered to be
too separated for good phase connection at 1.4 GHz; but this was not
known at the beginning of the project.  In addition, we hoped to rely
on the next stage, in-beam calibration, for more accurate imaging.

    Images for the four pulsars and their in-beam calibrators were
made after phase connection to their nodding calibrator.  As
summarized in Table 1, we did detect B0919+06 and B1857-26 and their
in-beam calibrator, but failed to detect the B1822-09 pulsar or its
in-beam calibrator.  The pulsar B0950+08 was easily detected, but its
in-beam calibrator was not.  The images of the detected pulsars and
in-beam calibrators were significantly distorted because of the large
angular distance between the nodding calibrator and the pulsar field.
We estimate that the position accuracy was no better than about 10 mas
and most images showed several secondary peaks.

    For further analysis of the strong pulsar B0950+08, where there is
no in-beam calibrator available for further phase calibrations, other
methods are being investigated to determine and remove the residual
ionospheric phase errors and will be reported on elsewhere
(\cite{bri99}).  More specifically, the dependence of pulsar position
(or visibility phase) with frequency can be used to determine the
amount of ionospheric refraction (the cause for the image distortions)
which can then be removed to produce an improved image.

\subsection{Using the In-beam Calibrator}

    For the pulsars B0119+06 and B1857-26 where both the pulsars and
in-beam calibrators were detected, we proceeded with the next stage of
phase connection between the in-beam calibrator and the pulsar (or
vice-versa).  Since the in-beam calibrator and/or pulsar are likely to
be relatively weak, special considerations are needed, especially
those needed to increase the coherence time in order to determine the
phase calibration with small errors.  For this reason the following
section will be somewhat detailed and associated with the AIPS
reduction package, generally used for VLBA calibrations.

    Choose the stronger of the in-beam calibrator or the pulsar (it
may be gated in order to increase the signal to noise) as the primary
phase reference.  If the correlation position of this source is not
within about 50 mas of the true source position, shift the phase
center of the data appropriately.  Otherwise, phase drifts due to this
large position error will decrease the effective coherence time of the
data and produce phases differences between the individual frequency
channels.  Since we are dealing with weak sources, a more accurate
phase solution can be made with longer integration times, or with
combined frequency channels.

    The AIPS calibration program, CALIB, determines the calibration
phase $\phi$ as a function of time $t$, frequency $\nu$ and telescope
$i$, $\phi(t, \nu, i)$, for the in-beam calibrator.  This program
essentially determines the calibration phase needed in order to
produce a point source from the existing data.  Since the data have
already been calibrated with respect to the nodding calibrator (and an
image, even if distorted, already made), the additional phase
calibrations should not be very variable in time, permitting the
averaging of the data for many minutes.  For
this reason the initial use of the nodding calibration is important to
increase the coherence time of the data associated with the in-beam
calibrator in order to use weaker sources.

    Before running CALIB some consideration of the expected signal to
noise of the solutions should be determined.  For VLBA observations
the nominal rms noise associated with a phase calibration solution,
using one minute of integration, with 8 MHz bandwidth, is 20 mJy.
\footnote{See VLBA sensitivities on NRAO web site.  For all subsequent
calculations we will assume that the observations used the entire VLBA
at 1.5 GHz, system temperature of 40K, with 8 recorded frequencies
each with bandwidth of 8 MHz.  This sensitivity should be scaled by
the relative sensitivity of the array, the number of telescopes used
for the self-calibration solution, the integration time and the total
bandwidth used in the solution.}
For example, if the correlated flux density of the source is $>50$
mJy, then the phase solution for each frequency channel of 8 MHz
bandwidth (there are eight of them) with one minute integration will
have signal to noise of about 2.5 to 1 which will produce an rms phase
error of about $20^\circ$.  Since this phase error is nearly
stochastic, the averaging of eight frequency channels over long
periods will average out these fluctations.

    For relatively weak in-beam calibrators, solution times of many
minutes and averaging of the frequency channels are required to obtain
valid solutions.  As another example, a source with 5 mJy correlated
flux density will require sufficient averaging to obtain a $<2.5$ mJy
noise level for each phase integration.  This would require a
integration time of 8 minutes and averaging of all 8 frequency
channels, each of 8 MHz, assuming the use of the VLBA.  While much
longer integration times and frequency averaging may increase the
signal to noise of the solutions, coherence may be lost.

    An illustration of the phase determination from a weak source is
shown in Table 2.  We have used the in-beam calibrator J1900-2602 for
the pulsar B1857-27 for the 1996 data.  For a range of parameters, we
have taken the phase calibration determined from J1900 and applied it
to B1857 which was then imaged.  Because this calibration method
determines those phases which make J1900 a point source, the image
quality obtained for J1900 {\it is no longer relevant}.  If the phase
calibration is sound, then the image of B1857 should display a
reasonable point source.

  The averaging time for the solutions ranged from 1 to 20 minutes
with a solution made for EACH frequency channel, or for ALL frequency
channels combined for better sensitivity.  In all cases the resultant
in-beam source looked point-like and its peak flux density was
approximately equal to the expected solution noise per averaging time.
In other words the phase determination algorithm does produce a point
source even with noise data.  However, when this phase calibration is
transferred to the pulsar and images are made, the peak flux density
and the quality of the pulsar images indicate the accuracy of the
phase calibration.  The solutions with at least five minutes solution
time, with all frequencies added together, produced good pulsar
images.  The slight decrease in pulsar peak flux from 5-min
integration to 20-min integration may be caused by loss of coherence
over this relatively long period of time.  The images in which all
frequencies have been averaged with a solution interval of 5, 10 or 20
minutes are acceptable and do not differ in the location of the peak
(See Table 2).

\section{THE RADIO IMAGES AND ASTROMETRIC RESULTS}

    The procedure outlined above was used, with a calibration solution
interval of five-minutes with all eight frequencies averaged together.
The peak flux density, and its spread over the three observations, of
the in-beam calibrator for B0919+06 and for B0857-26 are given in
Table 1.  Both were substantially unresolved.  A typical
calibrated phase solution is shown in Figure 1 for the 1996
observations of J1900-2602, the in-beam calibrator for B1857-26.

     After these in-beam calibration phases were applied to the pulsar
data, images were made for each of the three epochs.  They all showed
essentially a point source.  The images were then CLEANED with tight
boxes.  This process does increase the positional accuracy somewhat by
removing the distortions associated with the point-spread function and
permitting a better check on the quality of the image.  The position
of the pulsar was determined from a Gaussian-fit to the image.  The
position error is proportional to the resolution divided by the
signal-to-noise of the peak of the pulsar.

     Since the pulsar observations are tied to the same calibrator in
all three observations, they are on the same position grid and the
results from the three images can be directly compared to show the
motion of the pulsar.  Figure 2 shows the composite image for
B0919+06, where we have simply summed the three epoch images.  This
increases the noise background by a factor 1.7, but is illustrative of
the general results.

    Figure 3 shows the similar results for B1857-26.  Since the pulsar
was relatively weak during the 1994 observation (upper component) and
the location of the 1995 and 1996 positions were relatively close, we
did not use a simple sum of the three epochs to obtain this image.
This figure is composed of the representative parts of the images from
the three epochs of B1857-26, with no overlapping.  Both figures are
illustrative, with the proper motion and parallax fits made on the
positions derived from each observations.

    The results for all five epochs for B0919+06 are listed in Table
3.  With the additional two observations in 1998, a better analysis of
the accuracy of this experiment can be ascertained.  For simplicity we
have listed the relative position of B0916+06 for the five
observations; these relative positions are with respect to a nominal
position of B0916+06.  All positions have been tied to the in-beam
calibrator J0923+0638 with the assumed position given in the table.

    For B0919+06 the fit to the five epochs gives $\mu_\alpha= 17.7\pm
0.3$ mas/yr, $\mu_\delta = 79.2\pm 0.5$ mas/yr, $\pi = 0.31 \pm 0.14$
mas.  In Figure 4 the position of the pulsar for the five epochs,
after removal of the best-fit proper motion and position, is compared
with the parallax of 0.31 mas/yr, shown by the sinusoid.  The error
bars are those expected from the image noise and is equal to the image
resolution divided by the signal to noise at the peak of the pulsar.
Since three parameters have been determined (pulsar position, proper
motion and parallax) using four well-separated epochs, there is only
one degree of freedom, making the fit look better than it really is.
The agreement of the two 1998 observations, separated by four days, is
also better than expected.  The north/south motion of the pulsar is
much less sensitive to the parallax since the observations were
preferentially scheduled at maximum E/W parallax signal.  The N/S
scatter from the best position and proper motion is considerably
larger than that for the E/W direction.

     Our measured distance of B0919+06 is 3.2 (+2.6,-1.0) kpc and is
consistent with a limit of $>3$ kpc determined by \cite{tay93}.  The
previous estimate of this pulsar's proper motion is $\mu_\alpha= 13\pm
29$ mas/yr, $\mu_\delta=64\pm 37$ mas/yr, consistent with, but an
order of magnitude less accurate than, the present VLBA results
(\cite{har93}).

    The best fit proper motion and parallax for B1857-26, using just
three epochs, is $\mu_\alpha=-19.9\pm 0.3$ mas/yr, $\mu_\delta =
-47.3\pm 0.9$ max/yr, $\pi = 0.5 \pm 0.6$ mas.  The VLA results
(\cite{fom96}) gives $\mu_\alpha=-26\pm 5$ mas/yr, $\mu_\delta =
-47\pm 6$ mas/yr, in excellent agreement with the VLBA results.  The
distance limit derived with these observations of $>0.9$ kpc is
consistent with that derived by (\cite{tay93}) of 1.7 kpc.

\section{DISCUSSION}

    The parallax and proper motion obtained for B0919+06 and B1857-26
are among the most accurate yet obtained for pulsars or other galactic
objects (eg. \cite{bra99}).  The precision limits are consistent with
the signal to noise of the observations.  The additional epochs for
B0919+06 clearly improve the precision and decouple the proper motion
and parallax solutions, and we suggest that a minimum of five
well-separated epochs should be considered for obtaining accurate
parallaxes.  The consistency of the data with the fit for B0919+06
suggests that systematic errors at the level of 0.1 or 0.2 mas are not
significant when using in-beam calibrators within about $10'$ from the
target source.

    From analysis now underway on determining and removing the
ionospheric effects associated with B0950+08 (\cite{bri99}), we
estimate that the ionospheric refraction can lead to systematic error
of about 5 mas for a source-calibrator separation of about $7^\circ$.
Assuming that this systematic error is caused by the differential
ionospheric refraction between the calibrator and source, it should
decrease linearly with the source-calibrator separation since the
residual phase is a coherent difference rather than a stochastic
difference.  For a $12'$ separation we would expect such errors to be
about 0.15 mas in size.  This value is about the rms level of accuracy
of the present experiment.  With the additional epochs and the use of
pulsar gating, systematic errors (probably caused by differential
ionospheric refraction) may start to dominate the errors.  However,
removal of the ionospheric content by using the image changes over the
frequency range 1.4 to 1.7 GHz may reduce this error.

    When attempting to reach the 0.1 mas astrometric precision, the
variability in structure of the calibrator source can introduce
uncertainties at this level.  Many bright sources are 10 mas in size,
with variable core flux densities and moving components.  The cores
often shift with frequency because of optical depth effects.  Reaching
astrometric limits which are only 1\% of the source angular size can
be difficult.  Weaker calibrator sources, at the 10 mJy level, also
tend to be smaller in angular size since their stronger counterparts.
because of the $10^{12}$ K Compton-limit for extragalactic radio
sources.

     A general rule is that the closer the phase calibrator is to the
target source, the higher quality the images made with phase
referencing.  The strength of the calibrator is secondary as long as
it can be detected.  In other words, a source which is just barely
detected (say 3-sigma for a solution) will produce a phase error of
about 10 degrees which will be stochastic since it is determined from
random noise processes.  In contrast, the use of a very strong
calibrator, further away from the target, to obtain phase solutions
will have virtually no phase error component caused by noise, but
systematic phase errors of tens of degrees may persist over many
solution intervals and limit the dynamic range of the resulting images
of the target source.  Thus, the use of in-beam calibrators which are
weak generally provides better astrometric accuracy and image
fidelity.  These in-beam calibrators also provide simultaneous
calibration of the target source, whereas nodding calibrator phases
must be interpolated and then interpolated to apply to the target
source.

    The problem with routinely using in-beam calibrators is that the
field of view for which two sources can be simultaneously observed is
limited.  For the VLBA at 1.5 GHz, with a maximum separation of $25'$
of target and calibrator (both positioned somewhat within the
half-power circle of the primary beam), the NVSS Catalog
(\cite{con98}) contains an average of 20 sources above 2.5 mJy in such
an area, and about eight sources above 5.0 mJy.  At these flux density
levels, however, many sources may not have sufficient correlated flux
density to be detectable at 5000 km baselines.  With the present
sensitivity limits of the VLBA, a source should have about 5 mJy
correlated flux density to be detectable with 64 MHz bandwidth.
Weaker sources can be detected using the phased VLA with the VLBA with
128 MHz bandwidth.  If the pulsar is detectable, then an in-beam
calibrator is useful as the reference even with a correlated flux
density of 1 mJy.  Observations are now underway to search several
pulsar fields for possible in-beam calibrators and to determine the
proportion of faint mJy sources which are detectable with VLBI
resolution at 1.5 GHz.

\acknowledgments

The National Radio Astronomy Observatory is a facility of the National
Science Foundation operated under cooperative agreement by Associated
Universities, Inc.  This research at Cornell is supported by an NSF
grant AST 95-28391.

\clearpage

\begin{table}
\dummytable\label{tbl-1}
\begin{center}
\begin{tabular}{cccccl}
\multicolumn{6}{c}{TABLE 1} \\
\multicolumn{6}{c}{Observed Pulsars and Calibrators} \\
\tableline
Pulsar & Nodding & \hfil \hbox{Sep}\hfil & 
\hfil {In-beam}\hfil & \hfil \hbox{Sep}\hfil & Pk Flux  \\
{} & \hfil Cal \hfil & \hfil \hbox{deg}\hfil & \hfil Cal \hfil &
\hfil \hbox {min} & Density (mJy)\\
\tableline
B0919+06 & 0906+015 &   6 & J0923+0638 & 12 & $11.0\pm 0.5$\\
B0919+06 & J0914+0245&  4 & J0923+0638 & 12 & Used for last two epochs \\
B0950+08 & 1004+141 &   7 & J0948+0800 & 14 & In-beam not detected \\
B1822$-$09 & 1741$-$038 &  11 & J1825$-$0944 & 13 & Both not detected \\
B1857$-$26 & 1921$-$293 &   6 & J1900$-$2602 &  7 & $6.4\pm 0.4$\\
\tableline
\end{tabular}
\end{center}
\end{table}

\begin{table}
\dummytable\label{tbl-2}
\begin{center}
\begin{tabular}{clcccl}
\multicolumn{6}{c}{TABLE 2} \\
\multicolumn{6}{c}{Self-calibration Test on a 7-mJy Source} \\
\tableline
\hfil Solution \hfil & \hfil Freq \hfil & 
\hfil Solution \hfil & \hfil \hbox{In-Beam}\hfil & 
\multicolumn{2}{c}{Pulsar} \\
\hfil Inter (min) \hfil & \hfil Used \hfil &
\hfil Noise \hfil & \hfil \hbox{Pk Flux}\hfil &
\hfil \hbox {Pk Flux} \hfil & Quality \\
\tableline
 1   &   EACH & 20.0 &  17.9    &    1.2 & Noise\\
 3   &   EACH & 11.6 &  11.7    &    2.5 & Noise\\
 3   &   ALL  &  4.1 &   6.1    &    4.9 & Poor\\
 5   &   EACH &  8.9 &  10.1    &    3.0 & Poor\\
 5   &   ALL  &  3.2 &   5.7    &    6.1 & Good\\
10   &   EACH &  6.3 &   8.0    &    3.7 & Fair\\
10   &   ALL  &  2.2 &   4.8    &    5.6 & Good\\
20   &   EACH &  4.5 &   6.0    &    3.6 & Fair\\
20   &   ALL  &  1.6 &   3.9    &    5.0 & Good\\
\tableline
\end{tabular}
\end{center}
\end{table}

\begin{table}
\dummytable\label{tbl-3}
\begin{center}
\begin{tabular}{ccc}
\multicolumn{3}{c}{TABLE 3} \\
\multicolumn{3}{c}{Proper Motion for B0916+06} \\
\tableline
\hfil Observation \hfil & 
\multicolumn{1}{c}{E/W Offset*} & 
\multicolumn{1}{c}{N/S Offset*} \\
\hfil Date \hfil &
\multicolumn{1}{c} {(arcsec)} &
\multicolumn{1}{c} {(arcsec)}  \\
\tableline
1994.85 & $-0.0453\pm 0.0002$ & $-0.2640\pm 0.0004$ \\
1995.77 & $-0.0289\pm 0.0001$ & $-0.1892\pm 0.0002$ \\
1996.25 & $-0.0209\pm 0.0001$ & $-0.1437\pm 0.0002$ \\
1998.23 & $+0.0142\pm 0.0001$ & $+0.0076\pm 0.0002$ \\
1998.24 & $+0.0144\pm 0.0001$ & $+0.0087\pm 0.0002$ \\
\tableline
\multicolumn{3}{l}{* with respect to 09:22:14.000, +06:38:22.70}\\
\multicolumn{3}{l}{Assumed position of J0923+0638: 09:23:03.989, +06:38:58.98}\\
\end{tabular}
\end{center}
\end{table}

\clearpage

\figcaption[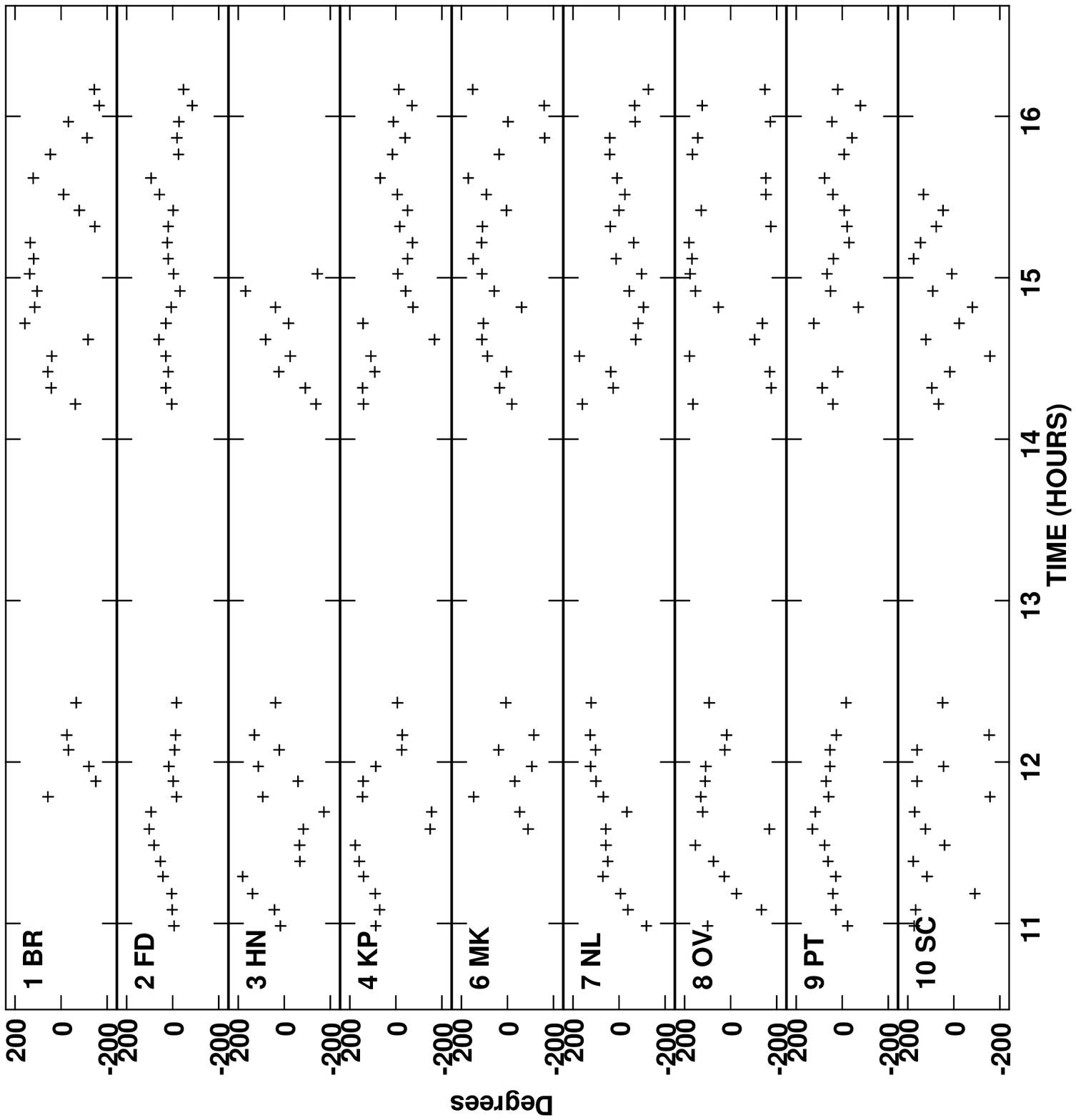]{The self-calibrator phases for the in-beam calibrator
J1900-2602:  Each plotted point shows the residual phase determined every
five minutes averaging all eight frequency channels.  Antenna 5 at Los
Alamos was the reference antenna.  The scatter becomes somewhat larger for
MK and SC, the longer baselines.  The continuity of the phases with time
is far from random for this 7-mJy source.}\label{fig1}

\figcaption[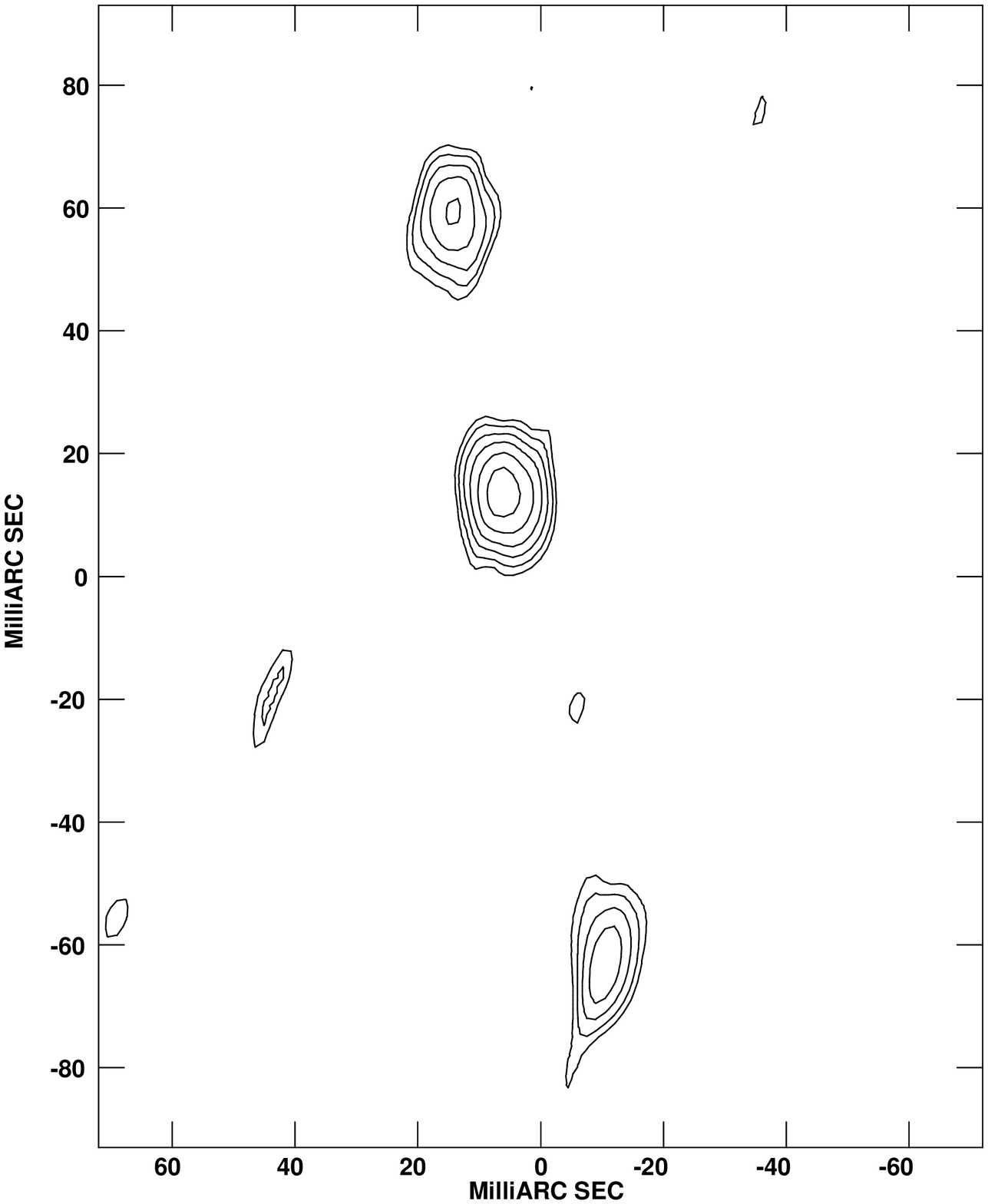]{The motion of Pulsar B0919+06: The sum of the
three images made for the pulsar B0919+06 after calibration with
J0923+0638, an 12 mJy in-beam calibrator.  The lowest contour level is
1.5 mJy with increasing steps of a factor 1.4.  The resolution is
$15\times 10$ mas in position angle 0.  The peak flux density at each
epoch is: (Bottom) 1994-4.7 mJy; (middle) 1995-10.0 mJy, (top)
1996-6.2 mJy.}\label{fig2}

\figcaption[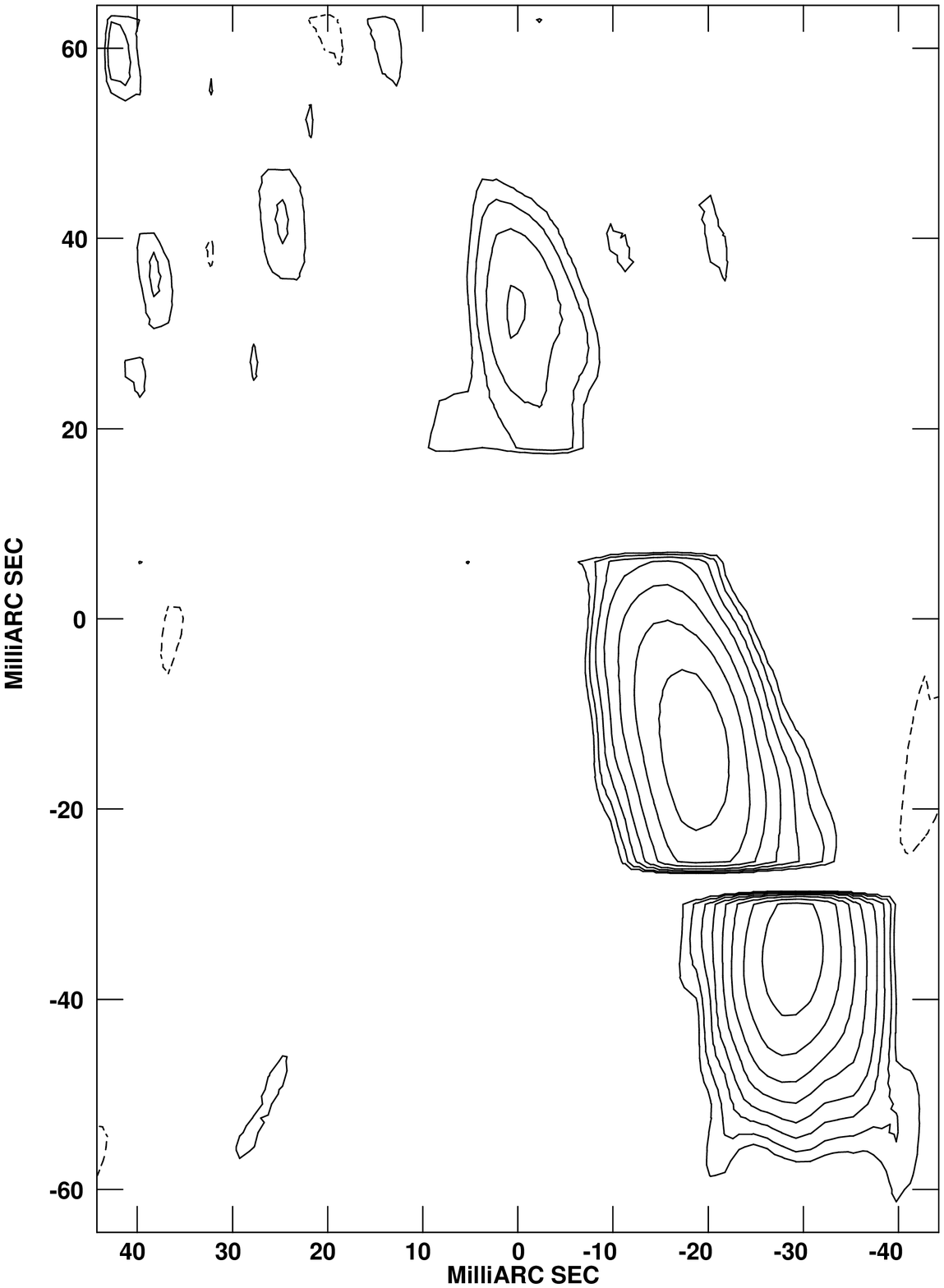]{The motion of Pulsar B1857-26: The composite of
the three images made for the pulsar B1857-26 after calibration with
J1900-2602, a 7 mJy in-beam calibrator.  The lowest contour level is
0.5 mJy with increasing steps of a factor 1.4.  The resolution is
$20\times 10$ mas in position angle 0.  The peak flux density at each
epoch is: (top) 1994-1.4 mJy; (middle) 1995-5.3 mJy, (bottom) 1996-6.9
mJy.}\label{fig3}

\figcaption[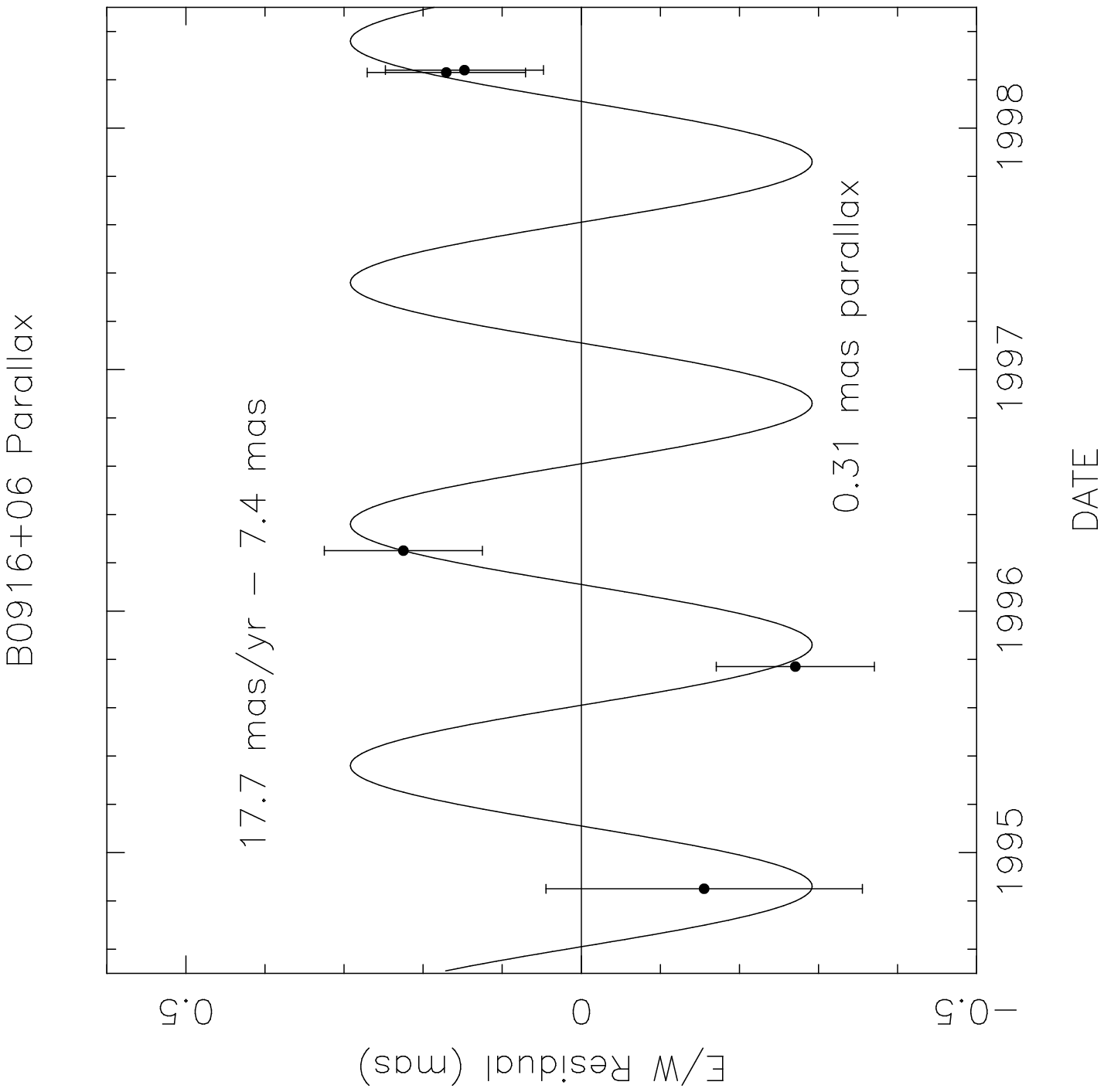]{The residual motion of Pulsar B0919+06: The comparison
of the measured E/W position of the pulsar after removal of the best
fit E/W position and proper motion listed in the plot.  The sinusoid
corresponds to that residual motion expected for a source with parallax
0.31 mas.}\label{fig4}

\clearpage
\begin{figure}
\plotone{fig1.eps}
\end{figure}

\clearpage
\begin{figure}
\plotone{fig2.eps}
\end{figure}

\clearpage
\begin{figure}
\plotone{fig3.eps}
\end{figure}

\clearpage
\begin{figure}
\plotone{fig4.eps}
\end{figure}

\end{document}